# Fast Link Recovery via PTP-synchronized Nanosecond Optical Switching

V. Yokar, A. Mehrpooya, Y. Teng, S. Shen, Z. Wu, K. Bardhi, S. Yan[*], D. Simeonidou

*HPN Group, Smart Internet Lab, University of Bristol, UK*

[*]*shuangyi.yan@bristol.ac.uk*

**Abstract:** This paper proposes and validates a PTP-synchronized 8.4ns optical switching with a 100ns jitter at the switching edges. This approach is adopted and demonstrated for instant network recovery within 2.7ms and scheduled network recovery.

## 1. Introduction

Google's recent deployment of MEMS-based optical circuit switching in data centres has inspired further explorations of optical switching in optical networks, promising more energy-efficient solutions [1]. While commercially available optical switches with nanosecond-level switching times have existed for decades, managing and configuring multiple optical switches becomes increasingly challenging as switching times reach microseconds or nanoseconds. Most solutions require both phase and frequency clock synchronization, along with dedicated paths, making them less scalable and robust [2]. Recent research has explored Precise Time Protocol (PTP) based time distribution and synchronization over Ethernet, which enables nanosecond-level synchronization across larger coverage areas using commercially available devices [3]. This has opened up the possibility of using PTP-based solutions to manage high-speed optical switches in optical networks.

In this paper, we proposed PTP-synchronised optical switching by configuring two 1×4 optical switches with 10 ns switching time simultaneously. The jitter of the two synchronized optical switches is measured within 100 ns over a half-hour period. As an application of PTP-synchronized optical switching, we demonstrated optical link recovery by detecting link failures and immediately switching to a backup link within 2.7 ms. The link can also be scheduled to recover at the designated time. To our knowledge, this is the first instance of integrating PTP-based time synchronization into network management.

## 2. PTP Synchronised Nanosecond-level Optical Switching

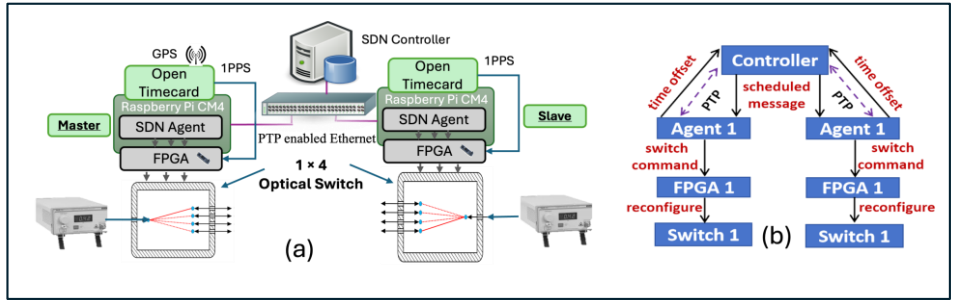

Fig. 1. Optical switching and control system, featuring master-slave GPS timing, FPGAs, optical controllers, and Open Timecard PTP-based switching; (b) SyncNet SDN controller hierarchy.

### 2.1. Open TimeCard driven ns-level optical switching with FPGA

As shown in Figure 1(a), the control system for the optical switches incorporates two Raspberry Pi Compute Module 4 (CM4) devices and two dedicated FPGAs. The Raspberry Pi acts as a local agent to receive both PTP synchronisation signals and SyncNet commands from the SDN controller. It is also equipped with an OpenTime card from Timebeat [4], which can provide accurate local time through GPS/GNSS. The OpenTime card generates a pulse-per-second signal to the FPGA. The FPGA serves as a high-speed switching driver for the 1x4 optical switch, enabling precise ns-level control. It receives a 1PPS signal from the grandmaster and slave device (OpenTime cards) and uses it to synchronize the switches with other components of the experiment. Additionally, the FPGA is connected to an SDN agent via a UART (Universal Asynchronous Receiver-Transmitter module) link, allowing for remote configuration changes. Based on the received commands, the FPGA can either modify the optical switch configuration in accordance with the 1PPS signal or switch states immediately based on the SDN controller commands, providing flexibility for dynamic experimental scenarios.

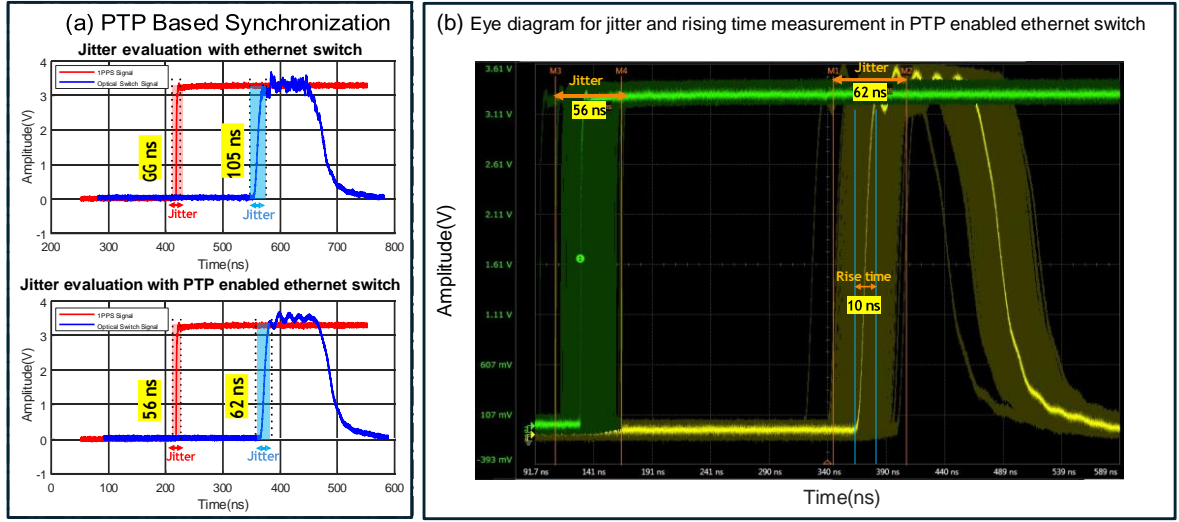

Fig. 2. Experimental results for (a) PTP-based synchronization through both Ethernet and PTP enabled switches (b) Jitter measurements over 1 half-hour period

### 2.2. SyncNet SDN controller

To manage multiple devices, we propose SyncNet to integrate PTP-based time synchronisation into the SDN controller. As shown in Figure 1(b), when a switch reconfiguration is required, the controller sends packets containing the target port and timestamp to each SDN agent. The SDN controller calculates its time offset with each agent using the PTP protocol and generates timestamps relative to their local clocks, eliminating the need for direct agent-to-agent synchronization. Once the local clock of the SDN agents reaches the designated timestamp, they send commands to the FPGAs to reconfigure the optical switches.

### 2.3. Synchronization validation

In Figure 1(a), the two optical switches are configured in a master-slave setup, to synchronize precise time for controlled 1×4 optical switches. The controller sends scheduled messages to agents that manage FPGA and optical switch configurations, enabling rapid and precise switching. To validate the optical switching with PTP-based synchronization, we set up experiments as each optical switch is configured to generate a 150-nanosecond switching window by switching two times. Two local agents of the optical switches are synchronised over a PTP-enabled Ethernet switch. The switching window of the slave optical switch is measured with a real-time oscilloscope, triggered by the master clock from the master switch. This setup enables us to evaluate the synchronization accuracy between two switches.

Figure 2(a) presents the jitter evaluation of the 1PPS signal and an emulated optical switch window under two different scenarios: (i) with a traditional ethernet switch and (ii) with a PTP-enabled ethernet switch. PTP-based synchronisation over the traditional ethernet switch exhibits a jitter of 105 ns. The relatively high jitter here reflects the limitations of standard Ethernet switches. By replacing the Ethernet switch with a PTP-enable ethernet switch, which gives priority to PTP traffic. In this case, the jitter reduces significantly to 56ns for the PPS signal and 62 ns for the optical switching window, representing a major improvement in timing precision. This reduction of approximately 40 per cent in jitter demonstrates the superior time synchronization capabilities of dedicated PTP-enabled switches compared to standard Ethernet switches. Figure 2(b) shows the long-term eye diagram of the switching window of the slave optical switch over a half-hour period. Based on this, we achieved PTP-based coordinated switching with a jitter of approximately 62 ns, while the rising edge of the pulse (switching time) is around 10 ns.

### 3. Fast Optical Link Recover with Synchronised Optical Switches

To evaluate PTP-synchronized optical switching, we deployed two optical switches in the lab testbed to explore optical connection recovery through a backup link as shown in Figure 3(a). The two optical links are connected via the optical switches. One Facebook Voyager Transponder emulates the original deployed network traffic using a 32 Gbaud PM-16QAM signal, while another provides optical traffic for the backup link, avoiding the transient effect of the EDFAs in the backup link.

Link failures are emulated by disconnecting the connections between fibre spools and the EDFA. The link power loss is detected using a photodiode, followed by signal processing in the FPGA. When the original link fails due to significant power loss, the optical signal must be switched to a backup link to restore the connection. To achieve recovery, both fast optical switches need to be configured simultaneously. The process requires the FPGA-based failure detection to notify the SDN controller, which then configures both switches for recovery. By combining PTP synchronization and SDN-controlled optical switching, the link can be recovered instantly or scheduled for recovery.

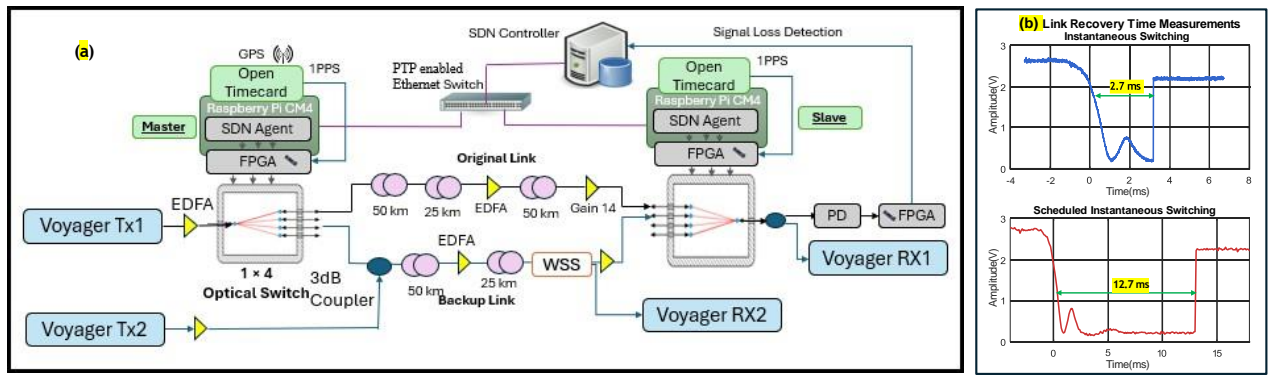

Fig. 3. (a) Experimental setup of the network recovery system, (b) Link recovery time measurements.

Figure 3(b) illustrates the link recovery time measurements across various switching scenarios: instantaneous link recovery and scheduled link recovery. For instantaneous recovery, the recovery time is measured at 2.7 ms, indicating the system's ability to detect power drop, notify the SDN controller, and recover from a failure within a very short timeframe. The designed solution also allows for scheduled recovery at any designated time, as measured the recovery time is around 12.7 ms in our experiment. It's important to note that the scheduled methods involve a 10 ms for the SDN controller and an additional 2.7 ms overhead for executing the actual switch. Although this approach offers pre-planned, deterministic behaviour, the extra overhead can be a limitation for applications needing faster recovery. In summary, while instantaneous recovery offers the fastest recovery for time-critical systems, scheduled recovery will reserve time for other components to prepare for the recovery process.

## 4. Conclusion

In this paper, we demonstrated PTP-synchronized optical switching with a jitter of 62ns and switching time of 10ns using commercially available components. The proposed solution integrates OpenTimeCard-based time synchronization with FPGA-controlled optical switches, enabling coordinated switching across multiple network nodes. We validated our approach through a network recovery demonstration, achieving instantaneous recovery within 2.7ms and scheduled recovery capabilities. The system's flexibility allows for both immediate response to failures and planned maintenance windows, making it suitable for various network scenarios. This work represents a significant step forward in applying PTP-based synchronization to optical network management, offering a scalable solution for future optical networks requiring precise timing control and rapid recovery capabilities.

## 5. Acknowledgements

This work was supported by the UK DSIT project REASON and EPSRC project (HASC, No. EP/X040569/1). Partial support was also provided by the CSA Catapult-UoB collaboration project. We extend our gratitude to Ian Gough and Lasse Johnsen from Timebeat for their contributions.